\documentclass[aps,prd,10pt,twocolumn,amsmath,superscriptaddress,amssymb,showpacs,floatfix,bibnotes,longbibliography]{revtex4-2}

\pdfoutput=1
\usepackage{hyperref}
\usepackage{graphicx}
\usepackage{amsfonts,amsmath,amssymb,bm,bbm,mathrsfs}
\usepackage{color}
\usepackage{url}
\usepackage{float}
\usepackage{slashed}
\usepackage{enumitem}
\usepackage{leftindex}
\usepackage[compat=1.1.0]{tikz-feynman}
\usepackage{feynmp-auto}
\usepackage[capitalise]{cleveref}
\usepackage{braket}

\newcommand{\TeV}{\mathrm{TeV}}
\newcommand{\keV}{\mathrm{keV}}
\newcommand{\cm}{\mathrm{cm}}

\hypersetup{
    pdfnewwindow=true,      % links in new window
    colorlinks=true,       % false: boxed links; true: colored links
    linkcolor=blue,          % color of internal links
    citecolor=blue,        % color of links to bibliography
    filecolor=blue,      % color of file links
    urlcolor=blue        % color of external links
}

\newcommand{\orcid}[1]{\begingroup
  \!\hypersetup{hidelinks}\href{https://orcid.org/#1}{\includegraphics[width=10pt]{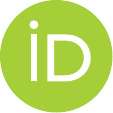}\!\!} \endgroup}

\newcommand{\appropto}{\mathrel{\vcenter{
  \offinterlineskip\halign{\hfil$##$\cr
    \propto\cr\noalign{\kern2pt}\sim\cr\noalign{\kern-2pt}}}}}

\begin{document}

\title{Inelastic Signatures of Electroweak Dark Matter}

\author{Juri Smirnov \orcid{0000-0002-3082-0929}\,}
\email{juri.smirnov@liverpool.ac.uk}
\affiliation{Department of Mathematical Sciences, 
\href{https://ror.org/04xs57h96}{University of Liverpool}, Liverpool, L69 7ZL, United Kingdom}

\author{Spencer~Griffith \orcid{0009-0002-1988-0768}\,}
\email{griffith.1037@osu.edu}
\affiliation{Center for Cosmology and AstroParticle Physics (CCAPP), \href{https://ror.org/00rs6vg23}{Ohio State University}, Columbus, OH 43210}
\affiliation{Department of Physics, \href{https://ror.org/00rs6vg23}{Ohio State University}, Columbus, OH 43210}

\author{John F. Beacom \orcid{0000-0002-0005-2631}\,}
\email{beacom.7@osu.edu}
\affiliation{Center for Cosmology and AstroParticle Physics (CCAPP), \href{https://ror.org/00rs6vg23}{Ohio State University}, Columbus, OH 43210}
\affiliation{Department of Physics, \href{https://ror.org/00rs6vg23}{Ohio State University}, Columbus, OH 43210}
\affiliation{Department of Astronomy, \href{https://ror.org/00rs6vg23}{Ohio State University}, Columbus, OH 43210}

\date{3 September 2026; revised 8 September 2026}

%%%%%%%%%%%%%%%%%%%%%%%%%%%%%%%%%%%%%%%%%%%%%%%%%%%%%%%%%
%%%%%%%%%%%%%%%%%%%%%%%%%%%%%%%%%%%%%%%%%%%%%%%%%%%%%%%%%

\begin{abstract}
Minimal dark matter extended to include a Majorana and a Dirac multiplet coupled through the Higgs --- the HC-MDM model --- provides a compelling and predictive framework.  We show that in certain limits of this model, the mass splitting and the inelastic interaction are representation-independent, while the thermal-relic masses are representation-dependent.  We then show that this model can account for the high-energy recoil event recently reported by LUX-ZEPLIN (LZ) while also respecting thermal-relic, elastic-scattering, and solar-capture constraints.  Generic inelastic DM models do not connect all these tests and generic Higgsino models do not survive them.
\end{abstract}

\maketitle

%%%%%%%%%%%%%%%%%%%%%%%%%%%%%%%%%%%%%%%%%%%%%%%%%%%%%%%%%
%%%%%%%%%%%%%%%%%%%%%%%%%%%%%%%%%%%%%%%%%%%%%%%%%%%%%%%%%

\textbf{\textit{Introduction.---}}
Minimal dark matter (MDM) is a compelling implementation of the weakly interacting massive particle solution for DM. In this model, a single $SU(2)_L$ multiplet is added to the Standard Model (SM), with all DM interactions arising from gauge invariance of the augmented Lagrangian. Consequently, the relic abundance and the direct-detection signatures are both fixed~\cite{Cirelli:2005uq, Cirelli:2007xd, Cirelli:2018iax}. We focus on fermion MDM, but scalar MDM is also possible.  For sufficiently heavy multiplets, nonperturbative Sommerfeld enhancement and bound-state formation complicate determination of the thermal-relic abundance, but these effects have been well examined~\cite{Hisano:2003ec, Hisano:2004ds, Hisano:2006nn, Arkani-Hamed:2008hhe, Cassel:2009wt, March-Russell:2008klu, vonHarling:2014kha, An:2016gad, Mitridate:2017izz}. 
Constraints from direct-detection experiments exclude pure Dirac electroweak multiplets because of tree-level elastic $Z$ exchange~\cite{Goodman:1984dc, CDMS:2005rss}, while odd (Majorana) multiplets up the 13-plet remain viable and detectable in next generation experiments~\cite{Bottaro:2021snn, Bloch:2024suj, Baudis:2024jnk, PANDA-X:2024dlo}. 

While MDM is compelling due to its simplicity, it omits interactions with the Higgs.  While this is acceptable, it does beg the question: \textit{Is there a minimal way to bring all interactions of the electroweak sector into play?}  This can be accomplished by adding both a Dirac (D) and a Majorana (M) multiplet to the SM. If the Majorana and Dirac multiplets are of adjacent size (e.g., $3_M 2_D$), then a Higgs coupling between the two multiplets is permitted. Such Higgs-coupled MDM (HC-MDM) models have been studied in particular cases (see, e.g., Refs.~\cite{Mahbubani:2005pt, DEramo:2007anh, Enberg:2007rp, Cohen:2011ec, Cheung:2013dua, Calibbi:2015nha, Freitas:2015hsa, Banerjee:2016hsk,Dedes:2014hga, Freitas:2015hsa, Beneke:2016jpw,Tait:2016qbg}) and more generally (see, e.g., Ref.~\cite{LopezHonorez:2017zrd}).  Then, in Ref.~\cite{GriffithSmirnov2026}, we showed that the $3_M2_D$, $5_M4_D$, $7_M6_D$, $9_M8_D$, $11_M10_D$, and $13_M12_D$ cases admit thermal-relic solutions over a broad mass range, $\sim$1--100~TeV, where we accounted for non-perturbative effects.  At a custodial point (discussed below), the tree-level Higgs coupling of the lightest neutral state to nucleons vanishes and the model is viable under current direct-detection constraints.

The recent LUX-ZEPLIN (LZ) search in an extended nuclear-recoil energy range allows a qualitatively new probe of HC-MDM models. Using a $2.84$ ton-year exposure and recoil energies up to $270~\keV$, LZ reported an event at $248\pm 23_{\rm stat}\pm 23_{\rm sys}~\keV$~\cite{LZ:2026axp}. The maximum local significance across the tested elastic and inelastic effective-field-theory models reaches $3.4\sigma$, with a global significance of $2.6\sigma$.  While far from a DM discovery, the expected backgrounds are low, making this event intriguing. Even more so because a standard elastic spin-independent interpretation is excluded, as the nuclear form factor suppresses high-energy events.

%%%%%%%%%%%%%%%%%%%%%%%%%%%%%%%%%%
\begin{figure}[t]
    \centering
    \includegraphics[width=0.98\columnwidth]{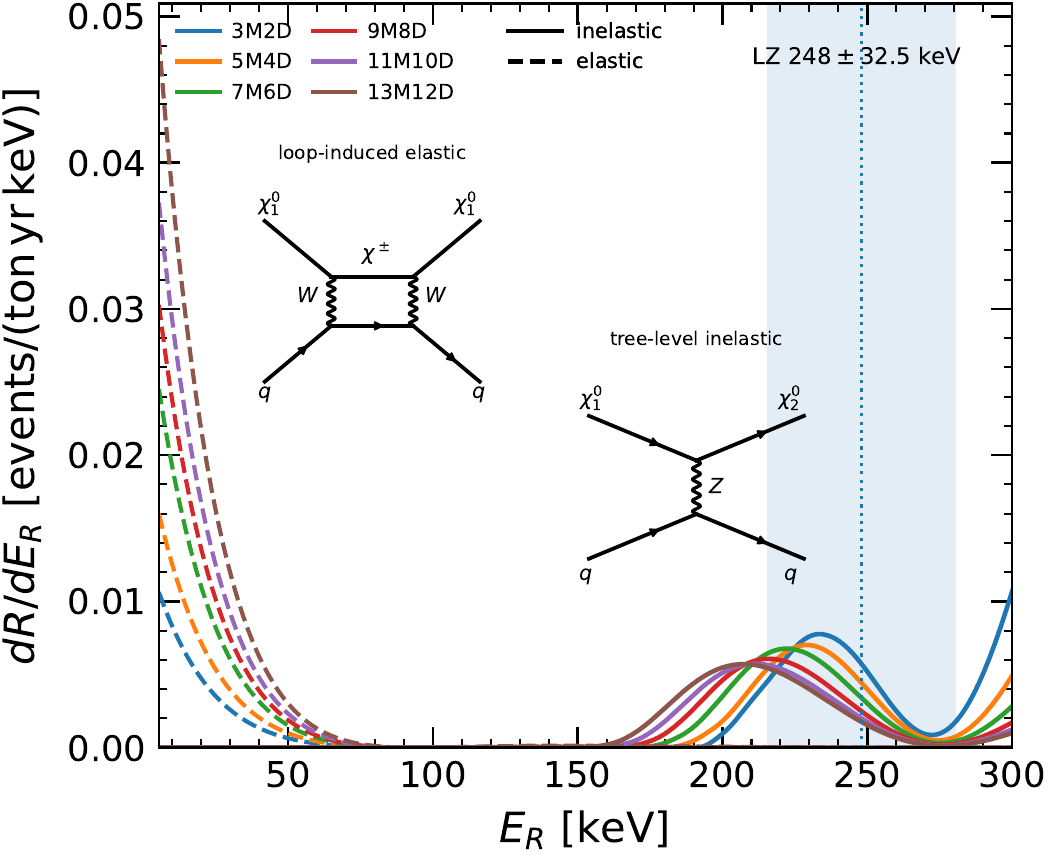}
    \caption{Recoil spectra for our benchmark models.  For each multiplet, the solid line is the tree-level inelastic contribution, while the dashed line is the loop-induced elastic contribution.  The LZ observation is shown by the blue band.}
    \label{fig:spectra}
\end{figure}
%%%%%%%%%%%%%%%%%%%%%%%%%%%%%%%%%%

A variety of new-physics interpretations are already being considered~\cite{Su:2026rwz, Fan:2026kxx, Freese:2026sga, Wu:2026nhi, Lou:2026idn, Nomura:2026qyq, Yin:2026jnn, DiMauro:2026ldr, Pospelov:2026ewn, Visinelli:2026kgt, Yamashita:2026ump, Du:2026guj, Rodd:2026tyn, McCabe:2026crm, Jeesun:2026vzo, Unwin:2026rdp, Dent:2026bji, deLima:2026shq, Gu:2026vto}.  A common approach is to invoke endothermic scattering, which can produce high-energy inelastic events while suppressing low-energy elastic scattering.  This has been emphasized for a nearly pure Higgsino, whose off-diagonal $Z^0$ coupling produces inelastic scattering between two neutral Majorana states with a splitting of a few hundred keV~\cite{Fan:2026kxx, Freese:2026sga, Wu:2026nhi, Yin:2026jnn, DiMauro:2026ldr}.  Very recently, Ref.~\cite{Pospelov:2026ewn} pointed out that such models are strongly limited by solar DM capture, though this can be relaxed if there is a substantial high-velocity DM population, as suggested in Refs.~\cite{Besla:2019xbx,Smith-Orlik:2023kyl}.

We show here that the same inelastic mechanism is present more generally in HC-MDM. The important additional feature in HC-MDM is that the Yukawa coupling that generates the neutral-state splitting also controls the thermal relic prediction. The observed high recoil energy suggests an inelastic splitting near the right scale. Because the $Z^0$-mediated interaction strength is fixed, we determine the splitting more precisely by matching the expected inelastic rate to the single observed LZ event. Combining the resulting universal (representation-independent) rate-matched relation with the thermal relic trajectory of HC-MDM then selects the DM mass and Yukawa coupling. Finally, given the mass and Yukawa strength, one can determine the size of the HC-MDM multiplets, which gives the loop-induced elastic scattering cross section.  Figure~\ref{fig:spectra} previews the predicted direct-detection signatures. 

%%%%%%%%%%%%%%%%%%%%%%%%%%%%%%%%%%%%%%%%%%%%%%%%%%%%%%%%%
%%%%%%%%%%%%%%%%%%%%%%%%%%%%%%%%%%%%%%%%%%%%%%%%%%%%%%%%%

\textbf{\textit{Setup of the HC-MDM model.---}}
We consider the HC-MDM Lagrangian for fermionic DM,
\begin{equation}
    \begin{split}
        & {\cal L} \supset \overline{D}(i\slashed{\mathcal{D}} - m_D)D + \frac12 \overline{M}(i\slashed{\mathcal{D}}-m_M)M \\
        & -y_1 D M H^\ast -y_2 \overline{D} M H +\mathrm{h.c.},
    \end{split}
\label{eq:lagrangian}
\end{equation}
where $M$ is a Majorana multiplet with hypercharge $Y_M=0$ and $D$ is a Dirac multiplet with $Y_D=1/2$, dictated by the requirement that a multiplet have an electrically neutral component and the relation $Q=T_3+Y$.

We focus on the cases $3_M2_D$, $5_M4_D$, $7_M6_D$, $9_M8_D$, $11_M10_D$, and $13_M12_D$, ignoring the singlet-doublet case ($1_M2_D$) because its Majorana component is electroweak neutral, making the correlated elastic signal Higgsino-like and therefore qualitatively distinct from the MDM cases considered here.  

Following Ref.~\cite{GriffithSmirnov2026}, we work at the custodial point,
\begin{equation}
    m_M=m_D\equiv m_\chi,\qquad y_1=-y_2\equiv y.
\label{eq:custodial}
\end{equation}
At this point, the tree-level diagonal Higgs coupling of the lightest neutral state vanishes, which strongly suppresses ordinary Higgs-mediated elastic scattering, as required to conform to experimental constraints (see Refs.~\cite{Tait:2016qbg, LopezHonorez:2017zrd}).

HC-MDM may be viewed as a one-sided limit of the generalized electroweakino structure $(2n-1)_M\oplus(2n)_D\oplus(2n+1)_M$. At the custodial points, the vectorlike pair forms a $(2n,2)$ of $SU(2)_L\times SU(2)_R$, which after EWSB decomposes under the diagonal $SU(2)_V$ as $(2n-1)\oplus(2n+1)$~\cite{LopezHonorez:2017zrd}. For $n=1$ this corresponds to the bino--Higgsino--wino field content, while HC-MDM follows by decoupling one of the neighboring Majorana multiplets.

The thermal relic abundance nevertheless depends on $y$ because Higgs exchange changes the annihilation and bound-state formation rates in the early universe (before electroweak symmetry breaking), which is relevant for determining the relic abundance. For each multiplet pair, we therefore have an externally computed thermal trajectory
\begin{equation}
    m_\chi = m_{\rm th}^{(R_M,R_D)}(y),
\label{eq:mthermal}
\end{equation}
where $R_M$, $R_D$ are the representation sizes of the Majorana and Dirac multiplets. A more detailed discussion of the general HC-MDM model is provided in Refs.~\cite{LopezHonorez:2017zrd, GriffithSmirnov2026}. 

Note that the sub-MeV splittings generated after electroweak symmetry breaking are negligible compared with the freezeout temperature for the masses considered here, and we therefore use the thermal trajectories of Ref.~\cite{GriffithSmirnov2026} without modification.

%%%%%%%%%%%%%%%%%%%%%%%%%%%%%%%%%%%%%%%%%%%%%%%%%%%%%%%%%
%%%%%%%%%%%%%%%%%%%%%%%%%%%%%%%%%%%%%%%%%%%%%%%%%%%%%%%%%

\textbf{\textit{Neutral states and inelastic $\mathrm{Z^0}$ scattering.---}}
The direct-detection phenomenology relevant for the LZ high-energy event requires developing the theory after electroweak symmetry breaking. A key feature is that the neutral-sector phenomenology is independent of the multiplet size. For the cases considered here, the neutral Higgs coupling has a universal Clebsch--Gordan coefficient, $1/\sqrt{2}$; its derivation is given in the End Matter.

At the custodial point of Eq.~(\ref{eq:custodial}), we define
\begin{equation}
    D_\pm= \frac{\psi^0\pm\widetilde \psi^0}{\sqrt2},
\label{eq:Dpm}
\end{equation}
where $\psi^0$ and $\widetilde\psi^0$ are the neutral Weyl components of the Dirac multiplet with $T_3=\mp1/2$. In the $(M,D_-,D_+)$ basis, the neutral mass matrix becomes
\begin{equation}
    {\cal M}_0=
    \begin{pmatrix}
        m_\chi & yv/\sqrt2 & 0\\
        yv/\sqrt2 & -m_\chi & 0\\
        0 & 0 & m_\chi
    \end{pmatrix}_{(M,D_-,D_+)}.
\label{eq:massmatrixcustodial}
\end{equation}
Thus $D_+$ decouples with mass $m_\chi$, while the $(M,D_-)$ block has physical mass
\begin{equation}
    M_\ast = \sqrt{m_\chi^2+\frac12y^2v^2}.
\end{equation}
The neutral spectrum therefore consists of one state at $m_\chi$ and two degenerate states at $M_\ast$, with splitting
\begin{equation}
    \delta(m_\chi,y) = \sqrt{m_\chi^2+\frac12y^2v^2}-m_\chi \simeq \frac{y^2v^2}{4m_\chi}, \qquad yv\ll m_\chi .
\label{eq:splitting}
\end{equation}
This splitting is universal in the $3_M2_D$--$13_M12_D$ cases.

The neutral-current interaction is similarly universal. In the $(M,D_-,D_+)$ basis,
\begin{equation}
    {\cal L}_Z= -\frac{g}{2c_W}Z_\mu \left[D_+^\dagger\bar\sigma^\mu D_- + D_-^\dagger\bar\sigma^\mu D_+ \right].
\label{eq:universalZ}
\end{equation}
Defining the physical mass eigenstates by
\begin{equation}
    m_{\chi_1}=m_\chi, \qquad m_{\chi_2}=m_{\chi_3}=M_\ast=m_\chi+\delta,
\label{eq:chi_masses}
\end{equation}
with $\chi_1\equiv D_+$ and $\chi_{2,3}$ spanning the $(M,D_-)$ sector, both excited states are kinematically accessible with the same splitting. Unitarity of the neutral-state rotation then gives
\begin{equation}
    \sum_{a=2,3} |g_{\chi_1\chi_a Z}|^2 = \left(\frac{g}{2c_W}\right)^2,
\label{eq:inclusiveZstrength}
\end{equation}
as shown in the End Matter.

The tree-level direct-detection process is therefore
\begin{equation}
    \chi_1+N\rightarrow\chi_{2,3}+N,
\label{eq:inelproc}
\end{equation}
where $N$ denotes a nucleon and the two excited states being experimentally indistinguishable for the present purpose. The corresponding inclusive neutron cross section is representation-independent and equal to the pseudo-Dirac neutral-current value,
\begin{equation}
    \sigma_n^{\rm inel} \simeq \frac{G_F^2\mu_{\chi n}^2}{2\pi} \simeq 7\times10^{-39}\ {\rm cm^2},
\label{eq:sigmainel}
\end{equation}
where $\mu_{\chi n}=m_\chi m_n/(m_\chi+m_n)\simeq m_n$ for the heavy DM masses considered here. Hence, at fixed $(m_\chi,y)$, both the neutral-state splitting and the inclusive tree-level inelastic scattering strength are representation-independent.

%%%%%%%%%%%%%%%%%%%%%%%%%%%%%%%%%%%%%%%%%%%%%%%%%%%%%%%%%
%%%%%%%%%%%%%%%%%%%%%%%%%%%%%%%%%%%%%%%%%%%%%%%%%%%%%%%%%

\textbf{\textit{Inelastic xenon recoil spectrum.---}}
For endothermic scattering with splitting $\delta>0$, the minimum DM velocity required for a nuclear recoil of energy $E_R$ is
\begin{equation}
    v_{\rm min}(E_R)= \frac{1}{\sqrt{2m_AE_R}} \left(\frac{m_AE_R}{\mu_{\chi A}}+\delta \right),
\label{eq:vmin}
\end{equation}
where $m_A$ is the nuclear mass and $\mu_{\chi A}$ is the DM-nucleus reduced mass. Unlike elastic scattering, $v_{\rm min}$ has a minimum at finite recoil energy,
\begin{equation}
    E_R^\star= \frac{\mu_{\chi A}}{m_A}\delta\simeq\delta,
\label{eq:estar}
\end{equation}
where the approximate equality holds for $m_\chi\gg m_{\rm Xe}$.

Schematically, the recoil spectrum is
\begin{equation}
    \frac{dR}{dE_R} = \frac{\rho_\chi}{m_\chi} \sum_A f_A \frac{\sigma_A^0}{2\mu_{\chi A}^2} F_A^2(E_R)\, \eta\!\left(v_{\rm min}(E_R)\right),
\label{eq:rate}
\end{equation}
where $f_A$ is the natural xenon isotope abundance, $F_A(E_R)$ is the nuclear form factor, and $\eta(v_{\rm min})$ is the usual halo integral. For vector $Z^0$ exchange, the coherent weak charge is
\begin{equation}
    Q_V(A,Z)= (A-Z)-(1-4\sin^2\theta_W)Z,
\end{equation}
and the zero-momentum nucleus cross section is
\begin{equation}
    \sigma_A^0= \sigma_n^{\rm inel} \frac{\mu_{\chi A}^2}{\mu_{\chi n}^2} Q_V^2.
\label{eq:sigmaA}
\end{equation}

Equations~(\ref{eq:splitting}), (\ref{eq:universalZ}), and (\ref{eq:rate}) show that, at fixed $(m_\chi,y)$, the high-energy xenon signal is representation-independent.

Figure~\ref{fig:spectra} shows the resulting recoil spectra for our benchmarks. The solid curves denote the tree-level inelastic signal, while the dashed curves show the suppressed loop-induced elastic contribution. The inelastic spectra do not terminate at the upper edge of the LZ signal window; they begin to rise again above the xenon form-factor diffraction minimum. The continuation of this spectrum into the higher-energy LZ sideband therefore provides an additional consistency test of the interpretation~\cite{Rodd:2026tyn}. We do not include this region in the rate matching below because the corresponding nuclear-recoil acceptance efficiency
has not yet been made public.

%%%%%%%%%%%%%%%%%%%%%%%%%%%%%%%%%%%%%%%%%%%%%%%%%%%%%%%%%
%%%%%%%%%%%%%%%%%%%%%%%%%%%%%%%%%%%%%%%%%%%%%%%%%%%%%%%%%

\textbf{\textit{LZ selects a universal relation between $\mathrm{m_\chi, y}$.---}}
LZ reports a single event reconstructed at
\begin{equation}
E_R=248\pm23_{\rm stat}\pm23_{\rm sys}~{\rm keV}.
\end{equation}
Rather than identifying $\delta$ directly with the reconstructed recoil energy through Eq.~(\ref{eq:estar}), we determine the preferred splitting from the event rate. For each DM mass, we calculate
\begin{equation}
N_{\rm inel}(m_\chi,\delta)
=
{\cal E}_{\rm LZ}
\int_{5.4\,{\rm keV}}^{270\,{\rm keV}}
dE_R\,\frac{dR_{\rm inel}}{dE_R}(m_\chi,\delta),
\end{equation}
with ${\cal E}_{\rm LZ}=2.84~{\rm ton\,yr}$.
At this stage we do not fold in the full detector response or reconstruct the LZ likelihood, so this should be understood as a rate-level estimate.

For every value of $m_\chi$, we define the central LZ-selected splitting implicitly by
\begin{equation}
    N_{\rm inel}\left(m_\chi,\delta_{\rm LZ}(m_\chi)\right) = 1.
\label{eq:deltaLZrate}
\end{equation}
Unlike the simple approximation $\delta_{\rm LZ}\simeq E_R$, the rate-selected splitting is mildly mass-dependent. Over the mass range relevant for HC-MDM, the resulting central splitting is about $350$--$380~\keV$. The predicted recoil spectra populate the vicinity of the observed $248~\keV$ event because the energy of a measured recoil need not coincide with the mass splitting itself.

To illustrate the statistical uncertainty associated with observing only one event, we also construct a conservative rate band using the two-sided $90\%$ Poisson interval for the expected mean, assuming negligible backgrounds,
\begin{equation}
    0.05 < \mu < 4.70.
\label{eq:LZPoissonBand}
\end{equation}
For each $m_\chi$, the lower and upper boundaries of the splitting band are obtained by solving
\begin{align}
    N_{\rm inel}\left(m_\chi,\delta_{\rm LZ}^{-}(m_\chi)\right) & =4.70,\\
    N_{\rm inel}\left(m_\chi,\delta_{\rm LZ}^{+}(m_\chi)\right) & =0.05.
\label{eq:deltaLZband}
\end{align}
Because the inelastic rate decreases rapidly with increasing splitting, the larger allowed event rate corresponds to the lower boundary in $\delta$, and vice versa.

%%%%%%%%%%%%%%%%%%%%%%%%%%%%%%%%%%
\begin{figure}[t]
    \centering
    \includegraphics[width=0.98\columnwidth]{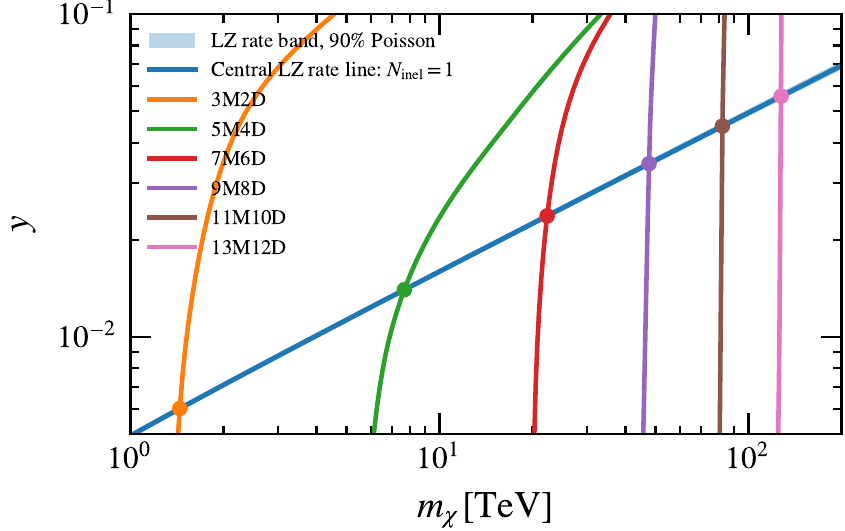}
    \caption{
    Universal $m_\chi$-$y$ relation selected by the LZ high-energy recoil. The central curve satisfies $N_{\rm inel}=1$, with the thin band obtained from the exact two-sided $90\%$ Poisson interval for one event. Representation-dependent thermal relic trajectories are superimposed; their intersections with the LZ band select the HC-MDM benchmark regions.
    }
    \label{fig:mass-y}
\end{figure}
%%%%%%%%%%%%%%%%%%%%%%%%%%%%%%%%%%

The mass-dependent splitting can then be mapped directly onto the Yukawa coupling using Eq.~(\ref{eq:splitting}). The corresponding upper and lower curves are obtained by replacing $\delta$ by $\delta_{\rm LZ}^{\pm}$. Because $\delta_{\rm LZ}\ll m_\chi$ in the parameter space of interest,
\begin{equation}
\label{eq:yLZ}
\end{equation}
The high-energy LZ event therefore selects a universal, rate-normalized band in the $(m_\chi,y)$ plane.

%%%%%%%%%%%%%%%%%%%%%%%%%%%%%%%%%%%%%%%%%%%%%%%%%%%%%%%%%
%%%%%%%%%%%%%%%%%%%%%%%%%%%%%%%%%%%%%%%%%%%%%%%%%%%%%%%%%

\textbf{\textit{Intersection with the thermal-relic trajectory.---}}
For a given multiplet pair, the relic abundance fixes
\begin{equation}
    m_\chi=m_{\rm th}^{(R_M,R_D)}(y),
\end{equation}
while the LZ recoil requires
\begin{equation}
\delta\!\left[m_{\rm th}^{(R_M,R_D)}(y),y\right]
=
\delta_{\rm LZ}\!\left(m_{\rm th}^{(R_M,R_D)}(y)\right).
\end{equation}
Combining the representation-independent LZ band with the
representation-dependent thermal-relic trajectories therefore selects
a model-dependent region
\begin{equation}
    \left\{y_\star,\,m_{\chi,\star}\right\}_{(R_M,R_D)} .
\end{equation}

Figure~\ref{fig:mass-y} illustrates the six intersections of the representation-dependent mass trajectories with the representation-independent value of the Yukawa coupling. Their locations are fixed jointly by the thermal freezeout dynamics and the LZ recoil rate. 

\begin{figure}[t]
    \centering
    \includegraphics[width=0.98\columnwidth]{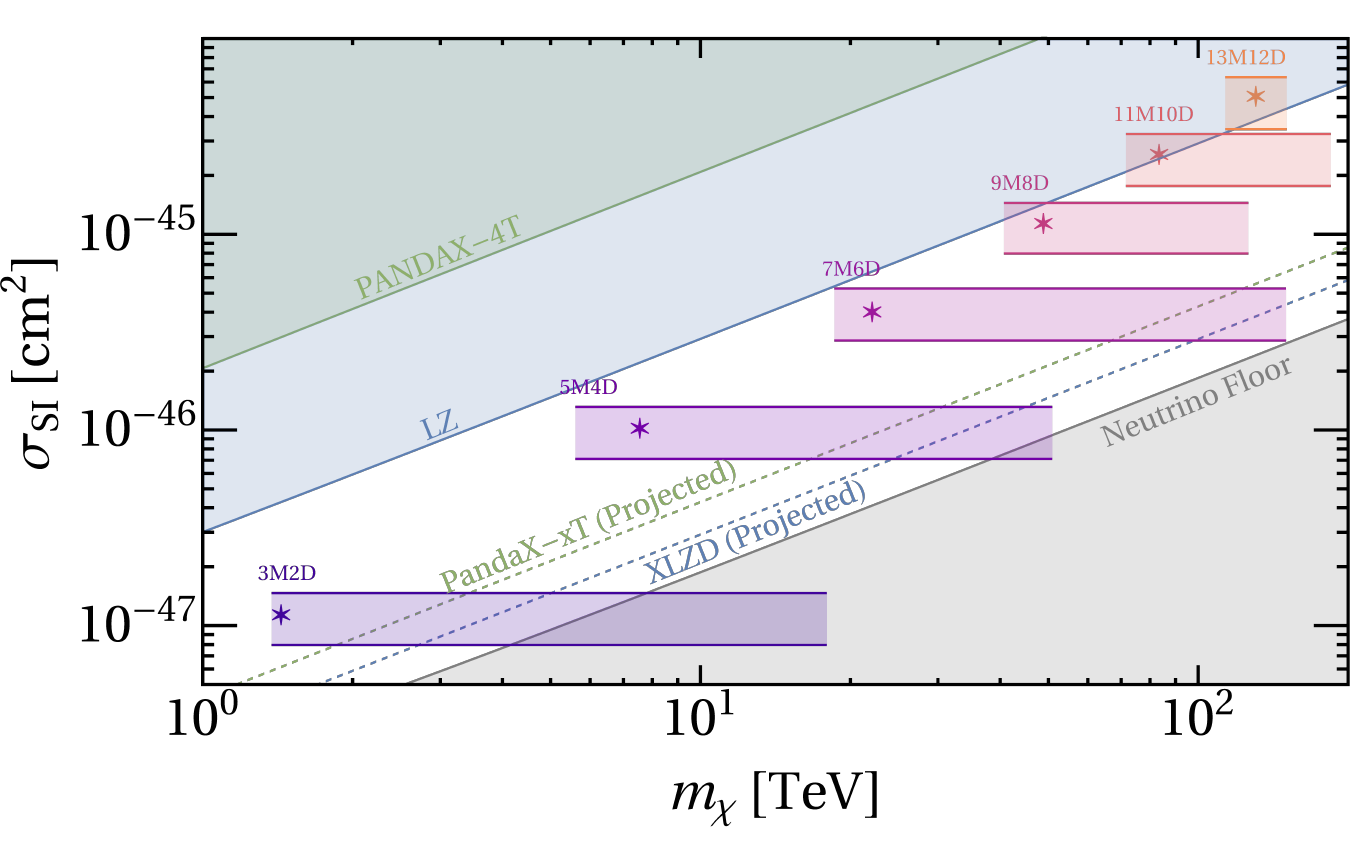}
    \caption{Thermal-relic mass ranges and spin-independent cross sections for the HC-MDM multiplets. The bars span the parameter space from a negligible $y$ (left) to the $s$-wave unitarity limit (right). The points selected by the LZ event (Fig.~\ref{fig:mass-y}) are indicated by stars.}
    \label{fig:mass ranges}
\end{figure}

Figure~\ref{fig:mass ranges} shows the corresponding thermal-relic mass ranges and spin-independent cross sections for each multiplet pair, as calculated in Ref.~\cite{GriffithSmirnov2026}. The stars denote our mass predictions for each multiplet based on the LZ observation, which predicts the expected elastic recoil rate and is one of the central results of this work.

%%%%%%%%%%%%%%%%%%%%%%%%%%%%%%%%%%
%%%%%%%%%%%%%%%%%%%%%%%%%%%%%%%%%%

%%%%%%%%%%%%%%%%%%%%%%%%%%%%%%%%%%%%%%%%%%%%%%%%%%%%%%%%%
%%%%%%%%%%%%%%%%%%%%%%%%%%%%%%%%%%%%%%%%%%%%%%%%%%%%%%%%%

\textbf{\textit{Correlated prediction for the elastic signal.---}}
At the custodial point, the tree-level Higgs-exchange contribution to spin-independent elastic scattering vanishes. The remaining signal is dominated by electroweak loops and is approximately the pure-MDM result,
\begin{equation}
    \sigma_{\rm SI}^{\rm loop} \simeq \sigma_{\rm SI}^{\rm MDM}(R_M).
\label{eq:sigloop}
\end{equation}
The mass dependence is weak for $m_\chi\gg m_N$, and the main hierarchy is set by the electroweak representation. This leads to a correlated two-channel prediction: a high-energy inelastic $Z^0$-mediated recoil together with a low-energy loop-induced elastic recoil.

Table~\ref{tab:benchmarks} shows our benchmark values for the LZ-selected points on each thermal trajectory including the expected event number from the loop-induced elastic cross section, shown by the stars in Fig.~\ref{fig:mass ranges}.  Here $N_{\rm el}$ denotes the expected number of raw loop-induced elastic recoil events evaluated over the same LZ exposure and recoil window as Eq.~(19). The corresponding inelastic yield is $N_{\rm inel}\simeq1$ for each benchmark by construction.

%%%%%%%%%%%%%%%%%%%%%%%%%%%%%%%%%%
\begin{table}[b]
\centering
\caption{Benchmark points selected by the intersections of the relic trajectories with the LZ rate-matched band.
The final column gives the corresponding expected elastic yield.}
\label{tab:benchmarks}
\setlength{\tabcolsep}{4pt}
\begin{tabular}{cc@{\hspace{3pt}}c@{\hspace{3pt}}ccc}
\hline\hline
Model
& $y_\star$
& \shortstack{$m_{\chi,\star}$\\$[\TeV]$}
& \shortstack{$\delta_\star$\\$[\keV]$}
& \shortstack{$\sigma_{\rm SI}^{\rm loop}$\\$[\cm^2]$}
& $N_{\rm el}$\\
\hline
$3_M2_D$
& $6.0\times10^{-3}$
& $1.4$
& $378$
& $1.3\times10^{-47}$
& $0.5$\\

$5_M4_D$
& $1.4\times10^{-2}$
& $7.7$
& $386$
& $1.1\times10^{-46}$
& $0.8$\\

$7_M6_D$
& $2.4\times10^{-2}$
& $22$
& $382$
& $4.5\times10^{-46}$
& $1.2$\\

$9_M8_D$
& $3.4\times10^{-2}$
& $48$
& $377$
& $1.2\times10^{-45}$
& $1.5$\\

$11_M10_D$
& $4.5\times10^{-2}$
& $82$
& $372$
& $2.6\times10^{-45}$
& $1.8$\\

$13_M12_D$
& $5.6\times10^{-2}$
& $130$
& $368$
& $5.0\times10^{-45}$
& $2.4$\\
\hline\hline
\end{tabular}
\end{table}
%%%%%%%%%%%%%%%%%%%%%%%%%%%%%%%%%%

%%%%%%%%%%%%%%%%%%%%%%%%%%%%%%%%%%%%%%%%%%%%%%%%%%%%%%%%%
%%%%%%%%%%%%%%%%%%%%%%%%%%%%%%%%%%%%%%%%%%%%%%%%%%%%%%%%%

\textbf{\textit{Phenomenological implications.---}}
How can varying interpretations of the high-energy LZ event --- from new physics to detector backgrounds --- be tested?  We first note three generic approaches.  Greater exposure in this recoil range with LZ and other detectors will probe events exactly like this.  Greater exposure at lower energies will test possible correlated elastic-scattering signals.  And extending the analysis range to even higher energies may provide new clues~\cite{Rodd:2026tyn}.  These basic approaches are the only ways to test, e.g., generic inelastic DM fits.  

However, the puzzle of the LZ event is most interesting for the most predictive models, as they can be tested in other ways.  The HC-MDM models we explore here are especially rich while also being broad enough that some cases may survive while others do not.  For direct-detection searches, the inelastic interaction probes the high-velocity tail of the DM velocity distribution through the threshold $v_{\min}(E_R,\delta)$, while the correlated elastic signal is controlled primarily by the electroweak representation.  Combining the two channels can therefore determine the multiplet structure while simultaneously probing the local DM velocity distribution.  Moreover, when the inelastic threshold lies close to the maximum accessible velocity, the rate has strong seasonal modulation~\cite{McCabe:2026crm}.

A complementary test arises from solar DM capture and annihilation producing high-energy neutrinos that can be probed by IceCube, as very recently shown by Ref.~\cite{Pospelov:2026ewn}.  As they note, this places decisive pressure on the $2_D1_M$ case, or equivalently simple Higgsino models~\cite{Freese:2026sga, Fan:2026kxx, Wu:2026nhi}.  For the HC-MDM models we consider, the results are more interesting (details are given in the End Matter).  For the $3_M2_D$ and $5_M4_D$ cases, we find that Standard Halo Model solutions are strongly constrained by solar capture, but can remain viable if the recoil originates from a fast velocity component. For the $7_M6_D$ case, the situation is borderline and benefits from the observed June timing. For the $9_M8_D$, $11_M10_D$, and $13_M12_D$ cases, we find that even the Standard Halo Model is allowed. There we also show that excited DM states will de-excite via photon emission, potentially providing an additional signature. 

%%%%%%%%%%%%%%%%%%%%%%%%%%%%%%%%%%%%%%%%%%%%%%%%%%%%%%%%%
%%%%%%%%%%%%%%%%%%%%%%%%%%%%%%%%%%%%%%%%%%%%%%%%%%%%%%%%%

\textbf{\textit{Conclusions.---}}
We have shown that Higgs-coupled minimal dark matter (HC-MDM) provides a natural inelastic interpretation of the LZ $248~\keV$ nuclear-recoil event~\cite{LZ:2026axp}. At the custodial point, Higgs-induced mixing produces a light neutral state and a degenerate excited neutral sector, separated by a splitting of a few hundred keV, while the $Z^0$ interaction connects the two sectors off diagonally. This naturally suppresses low-energy elastic scattering while allowing high-energy xenon recoils.

The key new feature is that the high-energy inelastic signal is independent of the multiplet, whereas thermal freezeout is not. The observed LZ recoil defines a universal band in the $(m_\chi,y)$ plane. For each multiplet pair, the observed relic abundance gives a different trajectory, $m_\chi=m_{\rm th}^{(R_M,R_D)}(y)$. The intersections fix $(m_\chi,y)$ up to the experimental uncertainty. The electroweak representation then predicts a correlated loop-induced elastic cross section in the usual low-energy recoil window.

If our interpretation is correct, future data will provide a network of correlated tests: the high-energy inelastic spectrum and low-energy elastic signal probe the DM mass, Higgs coupling, and electroweak representation, while their time dependence and consistency with solar capture probe the high-velocity structure of the local DM distribution. HC-MDM therefore turns a single possible inelastic recoil event into a simultaneous test of the particle physics and astrophysics of
dark matter.

%%%%%%%%%%%%%%%%%%%%%%%%%%%%%%%%%%%%%%%%%%%%%%%%%%%%%%%%%
%%%%%%%%%%%%%%%%%%%%%%%%%%%%%%%%%%%%%%%%%%%%%%%%%%%%%%%%%

\begin{acknowledgments}
JS was supported by the UK Research and Innovation Future Leader Fellowship MR/Y018656/1. SG and JFB were supported by National Science Foundation Grant Nos.\ PHY-2310018 and PHY-2609870. The authors used LLMs for assistance with language and editing. All scientific content and numerical results were independently developed and verified by the authors.
\end{acknowledgments}

%%%%%%%%%%%%%%%%%%%%%%%%%%%%%%%%%%%%%%%%%%%%%%%%%%%%%%%%%
%%%%%%%%%%%%%%%%%%%%%%%%%%%%%%%%%%%%%%%%%%%%%%%%%%%%%%%%%

\clearpage

\appendix

\vspace{1cm}
\centerline{\Large {\bf End Matter}}
\vspace{0.5cm}

%%%%%%%%%%%%%%%%%%%%%%%%%%%%%%%%%%%%%%%%%%%%%%%%%%%%%%%%%
%%%%%%%%%%%%%%%%%%%%%%%%%%%%%%%%%%%%%%%%%%%%%%%%%%%%%%%%%

\textbf{\textit{Neutral-state mixing and $Z^0$ coupling.---}}
Here we summarize the underlying representation-independent
neutral-sector structure. For a $(2n+1)_M(2n)_D$ model, the neutral Higgs interaction contains
\begin{equation}
    \left\langle j_D,-\frac12;\frac12,+\frac12 \middle|j_D+\frac12,0 \right\rangle = \frac{1}{\sqrt{2}},
\label{eq:neutralCGsupp}
\end{equation}
independent of $j_D$. The neutral Yukawa mixing is therefore universal throughout the $3_M2_D$--$13_M12_D$ cases.

After electroweak symmetry breaking, Eq.~(\ref{eq:lagrangian}) gives (see Ref.~\cite{LopezHonorez:2017zrd})
\begin{equation}
    {\cal M}_0=
    \begin{pmatrix}
        m_M & y_1v/2 & y_2v/2\\
        y_1v/2 & 0 & m_D\\
        y_2v/2 & m_D & 0
    \end{pmatrix}.
\label{eq:massmatrix}
\end{equation}
At the custodial point, defining $D_\pm=(\psi^0\pm\widetilde\psi^0)/\sqrt2$ and rotating to the $(M,D_-,D_+)$ basis gives Eq.~(\ref{eq:massmatrixcustodial}).  Thus $D_+$ is the light state, while the $(M,D_-)$ sector gives the two degenerate excited states of Eq.~(\ref{eq:chi_masses}).

The neutral current is
\begin{equation}
    {\cal L}_Z= \frac{g}{2c_W}Z_\mu \left(\widetilde\psi^{0\dagger}\bar\sigma^\mu\widetilde\psi^0 - \psi^{0\dagger}\bar\sigma^\mu\psi^0\right),
\end{equation}
which in the $D_\pm$ basis gives the off-diagonal interaction of Eq.~(\ref{eq:universalZ}). Writing
\begin{equation}
    D_+\equiv\chi_1,\qquad D_-\equiv U_{D_-2}\chi_2+U_{D_-3}\chi_3,
\end{equation}
unitarity implies
\begin{equation}
    |U_{D_-2}|^2+|U_{D_-3}|^2=1.
\label{eq:Dminusunitarity}
\end{equation}
Hence
\begin{equation}
    \sum_{a=2,3}|g_{\chi_1\chi_a Z}|^2 = \left(\frac{g}{2c_W}\right)^2,
\end{equation}
which establishes Eq.~(\ref{eq:inclusiveZstrength}) and the representation independence of the inclusive inelastic interaction.

%%%%%%%%%%%%%%%%%%%%%%%%%%%%%%%%%%%%%%%%%%%%%%%%%%%%%%%%%
%%%%%%%%%%%%%%%%%%%%%%%%%%%%%%%%%%%%%%%%%%%%%%%%%%%%%%%%%

\textbf{\textit{Solar constraints.---}}
The inelastic $Z^0$ interaction responsible for the xenon recoil event also leads to efficient DM capture in the Sun followed by annihilations that produce neutrinos, which can be probed by IceCube~\cite{IceCube:2025fcu}.  In particular, Ref.~\cite{Pospelov:2026ewn} finds that IceCube constraints applied to thermal Higgsino DM (which also applies to the lighter HC-MDM multiplets considered here) require $\delta \gtrsim 566~{\rm keV}$, as the solar capture rate rises rapidly for smaller $\delta$.  As we show below, this constrains the lower (and lighter) HC-MDM representations, while the higher (and heavier) ones are allowed because the solar-capture rate falls quadratically with increasing DM mass.

The first endothermic capture process,
\begin{equation}
    \chi_1 + A \to \chi_{2,3} + A,
\end{equation}
produces an excited state that de-excites via the radiative transition
\begin{equation}
    \chi_{2,3}\to \chi_1+\gamma.
\end{equation}
The analogous radiative de-excitation of a pseudo-Dirac Higgsino was computed in Ref.~\cite{KrallReece2018}, which finds
\begin{equation}
    \Gamma_\gamma =
    C^2\frac{e^2g^4}{256\pi^5}\frac{\delta^3}{m_\chi^2}.
    \label{eq:radiative-width}
\end{equation}
We evaluated the corresponding HC-MDM coefficient with \textsc{Package-X}~\cite{Patel:2015tea,Patel:2016fam}. In the pure-$D$ limit, the two gauge topologies, with the photon attached to the charged fermion or to the $W$, give $I_F = 2-2\pi\frac{m_W}{m_\chi}+ \mathcal{O}(m_W^2/m_\chi^2),\, I_W = 6-3\pi\frac{m_W}{m_\chi}+ \mathcal{O}(m_W^2/m_\chi^2),$ and enter Eq.~(\ref{eq:radiative-width}) through
\begin{equation}
    C_{\rm HC}=\frac{I_F+I_W}{4}
    =2-\frac{5\pi}{4}\frac{m_W}{m_\chi}+\ldots .
\end{equation}
For $j_D=n-\frac12$, the neutral state couples through $W^\pm$ to the singly charged components with ladder factors $|T_+|^2=n^2$ and $|T_-|^2=n^2-1$. Because the photon weights the two paths by their charges, the relevant group factor is $(+1)\,n^2+(-1)\,(n^2-1)=1,$ so the leading transition-dipole amplitude is independent of the multiplet size.  Thus
\begin{equation}
    \tau_\gamma \simeq
    0.90~{\rm s}
    \left(\frac{m_\chi}{7.63~{\rm TeV}}\right)^2
    \left(\frac{386~{\rm keV}}{\delta}\right)^3
    \left(\frac{2}{C_{\rm HC}}\right)^2,
\end{equation}
showing that the radiative transition is unsuppressed.  For the lower representations the de-excitation is fast compared with a solar passage; for the heavier multiplets the lifetime increases to the minute scale.  In principle, this could lead to new signatures where the de-excitation gamma ray becomes visible.  In addition, if the optical depth for DM scattering is appreciable, collisional de-excitation could change the capture calculation because these exothermic processes could undo marginal captures.

For the capture calculation, we use the \textsc{Asteria} framework of Ref.~\cite{Leane:2023woh}, using its single-scatter treatment and replacing the elastic recoil kernel by an endothermic one. Writing $w^2=u^2+v_{\rm esc}^2(r)$ and $\mu_{\chi A}=m_\chi m_A/(m_\chi+m_A)$, the local kernel is
\begin{equation}
    \begin{split}
        \Omega_A^-(w,r)={}&n_A(r)\,w\, \Theta\!\left(w^2-\frac{2\delta}{\mu_{\chi A}}\right) \int_{E_{\rm low}}^{E_+}dE_R\, \frac{d\sigma_A}{dE_R},\\
        E_\pm={}&\frac{\mu_{\chi A}^2}{2m_A} \left(w\pm\sqrt{w^2-\frac{2\delta}{\mu_{\chi A}}}\right)^2,\\
        E_{\rm low}={}& \max\!\left[E_-,\frac{1}{2}m_\chi u^2-\delta\right].
    \end{split}
\label{eq:inelastic-capture-kernel}
\end{equation}
The single-scatter capture rate then takes the standard form,
\begin{equation}
    \frac{dC_A}{dr}
    =4\pi r^2\frac{\rho_\chi}{m_\chi}
    \int du\,\frac{f(u)}{u}\,w\,\Omega_A^-(w,r).
\end{equation}
The second term in $E_{\rm low}$ implements the capture condition $E_R+\delta>\frac12m_\chi u^2$. This modification is used only for the optically thin single-scatter calculation, which is a good approximation in our case. 

The main remaining freedom is the high-velocity part of the local halo. For an observed recoil energy $E_R$, the minimum laboratory-frame speed is
\begin{equation}
    v_{\min}(E_R,\delta)=
    \frac{m_A E_R/\mu_{\chi A}+\delta}{\sqrt{2m_AE_R}} .
\end{equation}
For the $248~{\rm keV}$ event in xenon, splittings above $500~{\rm keV}$ require velocities well beyond the range of the Standard Halo Model. Numerically,
\begin{equation}
    \begin{split}
        3_M2_D:\qquad & v_{\min}(500~{\rm keV})\simeq 936~{\rm km\,s^{-1}},\\
        5_M4_D:\qquad & v_{\min}(500~{\rm keV})\simeq 915~{\rm km\,s^{-1}},
\end{split}
\end{equation}
with the required speed increasing by approximately $70$--$80~{\rm km\,s^{-1}}$ if $\delta$ is raised to $566~{\rm keV}$. Consequently, if the $3_M2_D$ or $5_M4_D$ interpretation is to evade the solar bound through a larger splitting, the event must originate from a non-standard fast component of the local distribution. A narrow component with speed $v_s$ contributes
\begin{equation}
    \eta_s(v_{\min}) \simeq \frac{f_s}{v_s}
    \Theta(v_s-v_{\min}),
\end{equation}
where $f_s$ is its local density fraction. Matching the observed event rate to the rate-selected Standard Halo Model solutions indicates that only a small fraction, parametrically $f_s\sim 10^{-4}$--$10^{-3}$ for $v_s\sim 0.9$--$1.0\times10^3~{\rm km\,s^{-1}}$, would be required. 

Interestingly, velocities of this order occur in cosmological Milky-Way--LMC simulations.  For example, Ref.~\cite{Smith-Orlik:2023kyl} finds that the combined MW+LMC halo integral extends beyond $v_{\min}\simeq950~{\rm km\,s^{-1}}$ for a present-day analog and reaches $v_{\min}\simeq1000~{\rm km\,s^{-1}}$ close to the LMC pericenter.  The local fraction of particles originating in the LMC is $\kappa_{\rm LMC}\simeq0.26\%$ in the present-day snapshot, while native MW particles are additionally accelerated by the dynamical response to the LMC.  Earlier tailored MW-LMC simulations similarly find an Earth-frame tail extending to $\sim900~{\rm km\,s^{-1}}$, with its largest velocities reached in June~\cite{Besla:2019xbx}.  Thus the required $f_s\sim10^{-4}$--$10^{-3}$ component is small compared with the overall sub-percent LMC-associated population seen in simulations, although the abundance specifically at $v\sim10^3~{\rm km\,s^{-1}}$ remains simulation-dependent. The relevance of this tail for endothermic scattering has been demonstrated explicitly in Ref.~\cite{Reynoso-Cordova:2024xqz}, where including the LMC extends xenon sensitivity towards larger inelastic splittings.

The $7_M6_D$ and higher cases are qualitatively different. For $7_M6_D$, we explicitly estimate the effect of the event timing by evaluating the recoil rate using the June laboratory-frame velocity distribution appropriate to the 16 June event ($v_{\rm lab}$ higher by $\simeq 12~{\rm km\,s^{-1}}$ than the time average). This shifts the rate-selected splitting from $\delta\simeq382~\keV$ in the time-averaged treatment to $\delta\simeq400~\keV$, which we use in Table~II. At this splitting the solar capture rate is already falling rapidly, rendering the $7_M6_D$ case borderline without requiring an additional halo component. For still larger representations, the solar capture rate is sufficiently suppressed that the standard time-averaged halo solutions are already compatible with the present estimate.

Table~\ref{tab:solar-benchmarks} provides a useful quantitative summary.  We quote $\Gamma_{WW}= C_\odot/2 \,{\rm Br}_{WW},$ assuming capture-annihilation equilibrium, with ${\rm Br}_{WW} = \mathcal{O}(1)$.  For the $7_M6_D$ case, we use the upper rate-compatible June benchmark, $\delta\simeq400~{\rm keV}$, while the other SHM entries use the time-averaged central LZ solutions.  For the two light representations we also show a common high-velocity benchmark, $\delta=570~{\rm keV}$, chosen to illustrate the splitting at which the solar rate is strongly reduced.  The corresponding $v_{\min}$ is evaluated at $E_R=248~{\rm keV}$.  For illustration, we also quote the high-$\delta$ annihilation rates obtained with the same diagnostic analytic solar profile used above. 

For the high-velocity benchmark, we take $\delta=570~{\rm keV}$. The corresponding Yukawa coupling is slightly larger than for the Standard Halo Model solution and therefore shifts the thermal relic mass. Solving approximately along the relic-density trajectories shown in Fig.~\ref{fig:mass-y}, for $3_M2_D$, we find $m_\chi \simeq 1.47~{\rm TeV}, \, y \simeq 7.45\times10^{-3}$, while for  $5_M4_D$, we find $m_\chi \simeq 8.5~{\rm TeV}, \, y \simeq 1.79\times10^{-2}$. Compared with the Standard Halo Model thermal solutions, the corresponding mass shifts are modest, at the level of approximately $2\%$ and $12\%$, respectively. We use these shifted masses when evaluating the illustrative $\delta=570~{\rm keV}$ solar-capture benchmarks.

In summary, the solar-capture constraint is an exciting new way to test HC-MDM models.  As detailed above, it constrains some, but not all, multiplets while also providing an avenue for probing the high-velocity component of the DM velocity distribution.

%%%%%%%%%%%%%%%%%%%%%%%%%%%%%%%%%%
\begin{table}[t]
\centering
\small
\setlength{\tabcolsep}{4.5pt}
\begin{tabular}{c|cc|ccc}
\hline
 & \multicolumn{2}{c|}{SHM benchmark}
 & \multicolumn{3}{c}{high-v benchmark}\\
Reps
& $\delta$
& $\Gamma_{WW}$
& $\delta$
& $v_{\min}$
& $\Gamma_{WW}$\\

& [keV]
& [$\mathrm{s}^{-1}$]
& [keV]
& [$\mathrm{km\,s^{-1}}$]
& [$\mathrm{s}^{-1}$]\\
\hline
$3_M2_D$   & 378 & $\sim 9 \times 10^{21}$ & 570 & 1020 & $\sim 10^{15}$\\
$5_M4_D$   & 386 & $\sim 3\times10^{20}$ & 570 & 1000 & $\sim 5\times10^{13}$\\
$7_M6_D$   & 400 & $\sim 10^{19}$ & --  & --  & --\\
$9_M8_D$   & 394 & $\sim 8 \times 10^{18}$ & --  & --  & --\\
$11_M10_D$ & 392 & $\sim  3\times10^{18}$ & --  & --  & --\\
$13_M12_D$ & 390 & $\sim  10^{18}$ & --  & --  & --\\
\hline
\end{tabular}
\vspace{0.2cm}
\caption{Illustrative solar $WW$ annihilation rates for the LZ-selected HC-MDM benchmarks.  The first pair of columns uses the Standard Halo Model (SHM), with the $7_M6_D$ entry evaluated at the June-favoured upper rate-compatible splitting.  The second set shows the two lower representations at a common $\delta=570~{\rm keV}$ high-velocity benchmark.  For orientation, the IceCube $WW$ annihilation-rate limit used in Ref.~\cite{Pospelov:2026ewn} at $m_\chi=1.08~{\rm TeV}$ is approximately $1.5\times10^{19}~{\rm s}^{-1}$.}
\label{tab:solar-benchmarks}
\end{table}

%%%%%%%%%%%%%%%%%%%%%%%%%%%%%%%%%%%%%%%%%%%%%%%%%%%%%%%%%
%%%%%%%%%%%%%%%%%%%%%%%%%%%%%%%%%%%%%%%%%%%%%%%%%%%%%%%%%

\clearpage
\twocolumngrid
\bibliography{bibliography}

@article{Cirelli:2005uq,
    author = "Cirelli, Marco and Fornengo, Nicolao and Strumia, Alessandro",
    title = "{Minimal dark matter}",
    eprint = "hep-ph/0512090",
    archivePrefix = "arXiv",
    reportNumber = "DFTT40-2005, IFUP-TH-2005-34",
    doi = "10.1016/j.nuclphysb.2006.07.012",
    journal = "Nucl. Phys. B",
    volume = "753",
    pages = "178--194",
    year = "2006"
}

@article{LopezHonorez:2017zrd,
    author = "Lopez Honorez, Laura and Tytgat, Michel H. G. and Tziveloglou, Pantelis and Zaldivar, Bryan",
    title = "{On Minimal Dark Matter coupled to the Higgs}",
    eprint = "1711.08619",
    archivePrefix = "arXiv",
    primaryClass = "hep-ph",
    reportNumber = "ULB-TH-17-21, LAPTH-058-17, ULB-TH/17-21, LAPTH-058/17",
    doi = "10.1007/JHEP04(2018)011",
    journal = "JHEP",
    volume = "04",
    pages = "011",
    year = "2018"
}

@article{Mitridate:2017izz,
    author = "Mitridate, Andrea and Redi, Michele and Smirnov, Juri and Strumia, Alessandro",
    title = "{Cosmological Implications of Dark Matter Bound States}",
    eprint = "1702.01141",
    archivePrefix = "arXiv",
    primaryClass = "hep-ph",
    reportNumber = "CERN-TH-2017-030, IFUP-TH-2017",
    doi = "10.1088/1475-7516/2017/05/006",
    journal = "JCAP",
    volume = "05",
    pages = "006",
    year = "2017"
}

@article{Bottaro:2021snn,
    author = "Bottaro, Salvatore and Buttazzo, Dario and Costa, Marco and Franceschini, Roberto and Panci, Paolo and Redigolo, Diego and Vittorio, Ludovico",
    title = "{Closing the window on WIMP Dark Matter}",
    eprint = "2107.09688",
    archivePrefix = "arXiv",
    primaryClass = "hep-ph",
    doi = "10.1140/epjc/s10052-021-09917-9",
    journal = "Eur. Phys. J. C",
    volume = "82",
    number = "1",
    pages = "31",
    year = "2022"
}

@article{Cohen:2011ec,
    author = "Cohen, Timothy and Kearney, John and Pierce, Aaron and Tucker-Smith, David",
    title = "{Singlet-Doublet Dark Matter}",
    eprint = "1109.2604",
    archivePrefix = "arXiv",
    primaryClass = "hep-ph",
    reportNumber = "MCTP-11-33, SLAC-PUB-14584",
    doi = "10.1103/PhysRevD.85.075003",
    journal = "Phys. Rev. D",
    volume = "85",
    pages = "075003",
    year = "2012"
}

@article{Cheung:2013dua,
    author = "Cheung, Clifford and Sanford, David",
    title = "{Simplified Models of Mixed Dark Matter}",
    eprint = "1311.5896",
    archivePrefix = "arXiv",
    primaryClass = "hep-ph",
    reportNumber = "CALT-68-2870",
    doi = "10.1088/1475-7516/2014/02/011",
    journal = "JCAP",
    volume = "02",
    pages = "011",
    year = "2014"
}

@article{Calibbi:2015nha,
    author = "Calibbi, Lorenzo and Mariotti, Alberto and Tziveloglou, Pantelis",
    title = "{Singlet-Doublet Model: Dark matter searches and LHC constraints}",
    eprint = "1505.03867",
    archivePrefix = "arXiv",
    primaryClass = "hep-ph",
    doi = "10.1007/JHEP10(2015)116",
    journal = "JHEP",
    volume = "10",
    pages = "116",
    year = "2015"
}

@article{March-Russell:2008klu,
    author = "March-Russell, John David and West, Stephen Mathew",
    title = "{WIMPonium and Boost Factors for Indirect Dark Matter Detection}",
    eprint = "0812.0559",
    archivePrefix = "arXiv",
    primaryClass = "astro-ph",
    reportNumber = "OUTP-08-01P",
    doi = "10.1016/j.physletb.2009.04.010",
    journal = "Phys. Lett. B",
    volume = "676",
    pages = "133--139",
    year = "2009"
}

@article{vonHarling:2014kha,
    author = "von Harling, Benedict and Petraki, Kalliopi",
    title = "{Bound-state formation for thermal relic dark matter and unitarity}",
    eprint = "1407.7874",
    archivePrefix = "arXiv",
    primaryClass = "hep-ph",
    reportNumber = "NIKHEF-2014-018",
    doi = "10.1088/1475-7516/2014/12/033",
    journal = "JCAP",
    volume = "12",
    pages = "033",
    year = "2014"
}

@article{An:2016gad,
    author = "An, Haipeng and Wise, Mark B. and Zhang, Yue",
    title = "{Effects of Bound States on Dark Matter Annihilation}",
    eprint = "1604.01776",
    archivePrefix = "arXiv",
    primaryClass = "hep-ph",
    reportNumber = "CALT-TH-2016-005",
    doi = "10.1103/PhysRevD.93.115020",
    journal = "Phys. Rev. D",
    volume = "93",
    number = "11",
    pages = "115020",
    year = "2016"
}

@article{Cassel:2009wt,
    author = "Cassel, S.",
    title = "{Sommerfeld factor for arbitrary partial wave processes}",
    eprint = "0903.5307",
    archivePrefix = "arXiv",
    primaryClass = "hep-ph",
    reportNumber = "OUTP-0910P",
    doi = "10.1088/0954-3899/37/10/105009",
    journal = "J. Phys. G",
    volume = "37",
    pages = "105009",
    year = "2010"
}

@article{Arkani-Hamed:2008hhe,
    author = "Arkani-Hamed, Nima and Finkbeiner, Douglas P. and Slatyer, Tracy R. and Weiner, Neal",
    title = "{A Theory of Dark Matter}",
    eprint = "0810.0713",
    archivePrefix = "arXiv",
    primaryClass = "hep-ph",
    doi = "10.1103/PhysRevD.79.015014",
    journal = "Phys. Rev. D",
    volume = "79",
    pages = "015014",
    year = "2009"
}

@article{Hisano:2003ec,
    author = "Hisano, Junji and Matsumoto, Shigeki and Nojiri, Mihoko M.",
    title = "{Explosive dark matter annihilation}",
    eprint = "hep-ph/0307216",
    archivePrefix = "arXiv",
    reportNumber = "ICRR-REPORT-500-2003-4, YITP-03-42",
    doi = "10.1103/PhysRevLett.92.031303",
    journal = "Phys. Rev. Lett.",
    volume = "92",
    pages = "031303",
    year = "2004"
}

@article{Hisano:2004ds,
    author = "Hisano, Junji and Matsumoto, Shigeki. and Nojiri, Mihoko M. and Saito, Osamu",
    title = "{Non-perturbative effect on dark matter annihilation and gamma ray signature from galactic center}",
    eprint = "hep-ph/0412403",
    archivePrefix = "arXiv",
    reportNumber = "ICRR-REPORT-513-2004-11, YITP-04-73",
    doi = "10.1103/PhysRevD.71.063528",
    journal = "Phys. Rev. D",
    volume = "71",
    pages = "063528",
    year = "2005"
}

@article{Hisano:2006nn,
    author = "Hisano, Junji and Matsumoto, Shigeki and Nagai, Minoru and Saito, Osamu and Senami, Masato",
    title = "{Non-perturbative effect on thermal relic abundance of dark matter}",
    eprint = "hep-ph/0610249",
    archivePrefix = "arXiv",
    reportNumber = "KEK-TH-1111",
    doi = "10.1016/j.physletb.2007.01.012",
    journal = "Phys. Lett. B",
    volume = "646",
    pages = "34--38",
    year = "2007"
}

@article{Cirelli:2018iax,
    author = "Cirelli, Marco and Gouttenoire, Yann and Petraki, Kalliopi and Sala, Filippo",
    title = "{Homeopathic Dark Matter, or how diluted heavy substances produce high energy cosmic rays}",
    eprint = "1811.03608",
    archivePrefix = "arXiv",
    primaryClass = "hep-ph",
    doi = "10.1088/1475-7516/2019/02/014",
    journal = "JCAP",
    volume = "02",
    pages = "014",
    year = "2019"
}

@article{Bloch:2024suj,
    author = "Bloch, Itay M. and Bottaro, Salvatore and Redigolo, Diego and Vittorio, Ludovico",
    title = "{Looking for WIMPs through the neutrino fogs}",
    eprint = "2410.02723",
    archivePrefix = "arXiv",
    primaryClass = "hep-ph",
    doi = "10.1007/JHEP08(2025)216",
    journal = "JHEP",
    volume = "08",
    pages = "216",
    year = "2025"
}

@article{Goodman:1984dc,
    author = "Goodman, Mark W. and Witten, Edward",
    editor = "Srednicki, M. A.",
    title = "{Detectability of Certain Dark Matter Candidates}",
    reportNumber = "Print-85-0030 (PRINCETON)",
    doi = "10.1103/PhysRevD.31.3059",
    journal = "Phys. Rev. D",
    volume = "31",
    pages = "3059",
    year = "1985"
}

@article{CDMS:2005rss,
    author = "Akerib, D. S. and others",
    collaboration = "CDMS",
    title = "{Limits on Spin-Independent Wimp-Nucleon Interactions from the Two-Tower Run of the Cryogenic Dark Matter Search}",
    eprint = "astro-ph/0509259",
    archivePrefix = "arXiv",
    reportNumber = "FERMILAB-PUB-05-455-E, SLAC-PUB-15992",
    doi = "10.1103/PhysRevLett.96.011302",
    journal = "Phys. Rev. Lett.",
    volume = "96",
    pages = "011302",
    year = "2006"
}

@article{Cirelli:2007xd,
    author = "Cirelli, Marco and Strumia, Alessandro and Tamburini, Matteo",
    title = "{Cosmology and Astrophysics of Minimal Dark Matter}",
    eprint = "0706.4071",
    archivePrefix = "arXiv",
    primaryClass = "hep-ph",
    reportNumber = "IFUP-TH-2007-12, SACLAY-T07-052",
    doi = "10.1016/j.nuclphysb.2007.07.023",
    journal = "Nucl. Phys. B",
    volume = "787",
    pages = "152--175",
    year = "2007"
}

@article{Baudis:2024jnk,
    author = "Baudis, Laura",
    title = "{DARWIN/XLZD: A future xenon observatory for dark matter and other rare interactions}",
    eprint = "2404.19524",
    archivePrefix = "arXiv",
    primaryClass = "astro-ph.IM",
    doi = "10.1016/j.nuclphysb.2024.116473",
    journal = "Nucl. Phys. B",
    volume = "1003",
    pages = "116473",
    year = "2024"
}

@article{PANDA-X:2024dlo,
    author = "Abdukerim, Abdusalam and others",
    collaboration = "PANDA-X, PandaX",
    title = "{PandaX-xT{\textemdash}A deep underground multi-ten-tonne liquid xenon observatory}",
    eprint = "2402.03596",
    archivePrefix = "arXiv",
    primaryClass = "hep-ex",
    doi = "10.1007/s11433-024-2539-y",
    journal = "Sci. China Phys. Mech. Astron.",
    volume = "68",
    number = "2",
    pages = "221011",
    year = "2025"
}

@article{Tait:2016qbg,
    author = "Tait, Tim M. P. and Yu, Zhao-Huan",
    title = "{Triplet-Quadruplet Dark Matter}",
    eprint = "1601.01354",
    archivePrefix = "arXiv",
    primaryClass = "hep-ph",
    reportNumber = "UCI-HEP-TR-2015-25",
    doi = "10.1007/JHEP03(2016)204",
    journal = "JHEP",
    volume = "03",
    pages = "204",
    year = "2016"
}

@article{Freitas:2015hsa,
    author = "Freitas, Ayres and Westhoff, Susanne and Zupan, Jure",
    title = "{Integrating in the Higgs Portal to Fermion Dark Matter}",
    eprint = "1506.04149",
    archivePrefix = "arXiv",
    primaryClass = "hep-ph",
    reportNumber = "PITT-PACC-1507",
    doi = "10.1007/JHEP09(2015)015",
    journal = "JHEP",
    volume = "09",
    pages = "015",
    year = "2015"
}

@article{Mahbubani:2005pt,
    author = "Mahbubani, Rakhi and Senatore, Leonardo",
    title = "{The Minimal model for dark matter and unification}",
    eprint = "hep-ph/0510064",
    archivePrefix = "arXiv",
    reportNumber = "MIT-CTP-3689, HUTP-05-A0044",
    doi = "10.1103/PhysRevD.73.043510",
    journal = "Phys. Rev. D",
    volume = "73",
    pages = "043510",
    year = "2006"
}

@article{DEramo:2007anh,
    author = "D'Eramo, Francesco",
    title = "{Dark matter and Higgs boson physics}",
    eprint = "0705.4493",
    archivePrefix = "arXiv",
    primaryClass = "hep-ph",
    doi = "10.1103/PhysRevD.76.083522",
    journal = "Phys. Rev. D",
    volume = "76",
    pages = "083522",
    year = "2007"
}

@article{Enberg:2007rp,
    author = "Enberg, R. and Fox, P. J. and Hall, L. J. and Papaioannou, A. Y. and Papucci, M.",
    title = "{LHC and dark matter signals of improved naturalness}",
    eprint = "0706.0918",
    archivePrefix = "arXiv",
    primaryClass = "hep-ph",
    reportNumber = "LBNL-62748, UCB-PTH-07-10",
    doi = "10.1088/1126-6708/2007/11/014",
    journal = "JHEP",
    volume = "11",
    pages = "014",
    year = "2007"
}

@article{Banerjee:2016hsk,
    author = "Banerjee, Shankha and Matsumoto, Shigeki and Mukaida, Kyohei and Tsai, Yue-Lin Sming",
    title = "{WIMP Dark Matter in a Well-Tempered Regime: A case study on Singlet-Doublets Fermionic WIMP}",
    eprint = "1603.07387",
    archivePrefix = "arXiv",
    primaryClass = "hep-ph",
    reportNumber = "IPMU16-0039, LAPTH-013-16",
    doi = "10.1007/JHEP11(2016)070",
    journal = "JHEP",
    volume = "11",
    pages = "070",
    year = "2016"
}

@article{Beneke:2016jpw,
    author = "Beneke, Martin and Bharucha, Aoife and Hryczuk, Andrzej and Recksiegel, Stefan and Ruiz-Femenia, Pedro",
    title = "{The last refuge of mixed wino-Higgsino dark matter}",
    eprint = "1611.00804",
    archivePrefix = "arXiv",
    primaryClass = "hep-ph",
    reportNumber = "TUM-HEP-1065-16, FTUAM-16-38, IFT-UAM-CSIC-16-106",
    doi = "10.1007/JHEP01(2017)002",
    journal = "JHEP",
    volume = "01",
    pages = "002",
    year = "2017"
}

@article{Dedes:2014hga,
    author = "Dedes, Athanasios and Karamitros, Dimitrios",
    title = "{Doublet-Triplet Fermionic Dark Matter}",
    eprint = "1403.7744",
    archivePrefix = "arXiv",
    primaryClass = "hep-ph",
    doi = "10.1103/PhysRevD.89.115002",
    journal = "Phys. Rev. D",
    volume = "89",
    number = "11",
    pages = "115002",
    year = "2014"
}

@article{GriffithSmirnov2026,
  author        = {Griffith, Spencer and Smirnov, Juri and Lopez-Honorez, Laura and Beacom, John F.},
  title         = {Minimal Dark Matter: Generalized Framework and Direct-Detection Sensitivity},
  eprint        = {2602.17764},
  archivePrefix = {arXiv},
  primaryClass  = {hep-ph},
  year          = {2026}
}

@article{LZ:2026axp,
    author = "Akerib, D. S. and others",
    collaboration = "LZ",
    title = "{Search for dark matter particle interactions in an extended nuclear recoil energy window with the LUX-ZEPLIN (LZ) experiment}",
    eprint = "2609.02823",
    archivePrefix = "arXiv",
    primaryClass = "hep-ex",
    month = "9",
    year = "2026"
}

@article{Su:2026rwz,
    author = "Su, Liangliang and Yang, Jin Min and Yang, Wen-Na",
    title = "{Inelastic Dark Matter Signature at High Recoil Energy in LUX-ZEPLIN and CRESST}",
    eprint = "2609.01475",
    archivePrefix = "arXiv",
    primaryClass = "hep-ph",
    month = "9",
    year = "2026"
}

@article{Fan:2026kxx,
    author = "Fan, JiJi and Reece, Matthew",
    title = "{Higgsino Above the Sea of Fog}",
    eprint = "2609.01504",
    archivePrefix = "arXiv",
    primaryClass = "hep-ph",
    month = "9",
    year = "2026"
}

@article{Freese:2026sga,
    author = "Freese, Katherine and Theodosopoulos, Dionysios P.",
    title = "{Higgsino Dark Matter Interpretation of the LUX-ZEPLIN 248 keV Nuclear-Recoil Event}",
    eprint = "2609.01583",
    archivePrefix = "arXiv",
    primaryClass = "hep-ph",
    month = "9",
    year = "2026"
}

@article{Wu:2026nhi,
    author = "Wu, Lei and Zhang, Yang and Zhu, Bin",
    title = "{TeV Higgsino Dark Matter from LZ Nuclear Recoil to Fermi-LAT Gamma Rays}",
    eprint = "2609.01590",
    archivePrefix = "arXiv",
    primaryClass = "hep-ph",
    month = "9",
    year = "2026"
}

@article{Lou:2026idn,
    author = "Lou, Yuanchao and Lu, Chih-Ting",
    title = "{Fermionic Dark Matter Absorption and the High-Energy Event in LUX-ZEPLIN}",
    eprint = "2609.01592",
    archivePrefix = "arXiv",
    primaryClass = "hep-ph",
    month = "9",
    year = "2026"
}

@article{Yin:2026jnn,
    author = "Yin, Wen",
    title = "{A PQ-Symmetric High-Scale SUSY Interpretation of the LZ High-Energy Recoil}",
    eprint = "2609.01892",
    archivePrefix = "arXiv",
    primaryClass = "hep-ph",
    month = "9",
    year = "2026"
}

@article{Nomura:2026qyq,
    author = "Nomura, Yasunori",
    title = "{Dark Matter as the Z{\_}2 Partner of the Standard Model Higgs Boson}",
    eprint = "2609.02505",
    archivePrefix = "arXiv",
    primaryClass = "hep-ph",
    reportNumber = "RIKEN-iTHEMS-Report-26",
    month = "9",
    year = "2026"
}

@article{DiMauro:2026ldr,
    author = "Di Mauro, Mattia",
    title = "{Dark Matter at the Kinematic Edge: Interpreting the 248 keV LZ Nuclear-Recoil Candidate}",
    eprint = "2609.02608",
    archivePrefix = "arXiv",
    primaryClass = "hep-ph",
    month = "9",
    year = "2026"
}

@article{Pospelov:2026ewn,
    author = "Pospelov, Maxim and Ramani, Harikrishnan",
    title = "{Strong Constraints on Higgsino Dark Matter from Solar Capture}",
    eprint = "2609.02775",
    archivePrefix = "arXiv",
    primaryClass = "hep-ph",
    month = "9",
    year = "2026"
}

@article{Visinelli:2026kgt,
    author = "Visinelli, Luca",
    title = "{A Peccei--Quinn Origin for Inelastic Electroweak Dark Matter after LUX-ZEPLIN}",
    eprint = "2609.02807",
    archivePrefix = "arXiv",
    primaryClass = "hep-ph",
    month = "9",
    year = "2026"
}

@article{Yamashita:2026ump,
    author = "Yamashita, Kimiko",
    title = "{Inelastic Dark Photon Dark Matter for the LUX-ZEPLIN High-Recoil Event and the Galactic Halo Gamma-Ray Excess}",
    eprint = "2609.02868",
    archivePrefix = "arXiv",
    primaryClass = "hep-ph",
    month = "9",
    year = "2026"
}

@article{Du:2026guj,
    author = "Du, Xiaokang and Wang, Fei",
    title = "{TeV Higgsino Interpretation of the LZ High-Recoil Event with Intermediate-Scale Electroweak Gauginos}",
    eprint = "2609.04163",
    archivePrefix = "arXiv",
    primaryClass = "hep-ph",
    month = "9",
    year = "2026"
}

@article{Rodd:2026tyn,
    author = "Rodd, Nicholas L. and Safdi, Benjamin R. and Slatyer, Tracy R. and Xu, Weishuang Linda",
    title = "{Confronting the Higgsino Interpretation of the LZ Event with the High-Energy Sideband}",
    eprint = "2609.04175",
    archivePrefix = "arXiv",
    primaryClass = "hep-ph",
    month = "9",
    year = "2026"
}

@article{McCabe:2026crm,
    author = "McCabe, Christopher",
    title = "{Seasonal dark matter from the LUX-ZEPLIN high-energy event}",
    eprint = "2609.04181",
    archivePrefix = "arXiv",
    primaryClass = "hep-ph",
    month = "9",
    year = "2026"
}

@article{Jeesun:2026vzo,
    author = "Jeesun, Sk and Majumdar, Anirban",
    title = "{Atmospheric neutrino up-scattering explanation of LZ 2026 excess}",
    eprint = "2609.04185",
    archivePrefix = "arXiv",
    primaryClass = "hep-ph",
    month = "9",
    year = "2026"
}

@article{Unwin:2026rdp,
    author = "Unwin, James",
    title = "{Axion Portal Dark Matter and the LUX-ZEPLIN High-Recoil Event}",
    eprint = "2609.04186",
    archivePrefix = "arXiv",
    primaryClass = "hep-ph",
    month = "9",
    year = "2026"
}

@article{Dent:2026bji,
    author = "Dent, James B. and Newstead, Jayden L.",
    title = "{Exothermic and Endothermic Inelastic Dark Matter Interpretations at LZ: Sideband Constraints and Future Prospects}",
    eprint = "2609.04673",
    archivePrefix = "arXiv",
    primaryClass = "hep-ph",
    month = "9",
    year = "2026"
}

@article{deLima:2026shq,
    author = "de Lima, Carlos Henrique",
    title = "{Exothermic Dark Matter at LZ}",
    eprint = "2609.05204",
    archivePrefix = "arXiv",
    primaryClass = "hep-ph",
    month = "9",
    year = "2026"
}

@article{Gu:2026vto,
    author = "Gu, Guanhua and Li, Lingfeng and Tang, Shao-Song and Xu, Yongheng",
    title = "{Inelastic from the Other Side: Xenon Excitation Signals in Light of the LZ High-Recoil Event}",
    eprint = "2609.05291",
    archivePrefix = "arXiv",
    primaryClass = "hep-ph",
    month = "9",
    year = "2026"
}

@article{IceCube:2025fcu,
    author = "Abbasi, R. and others",
    collaboration = "IceCube",
    title = "{Search for High-Energy Neutrinos From the Sun Using Ten Years of IceCube Data}",
    eprint = "2507.08457",
    archivePrefix = "arXiv",
    primaryClass = "hep-ex",
    month = "7",
    year = "2025"
}

@article{KrallReece2018,
author = {Krall, Rebecca and Reece, Matthew},
year = {2018},
month = {04},
pages = {043105},
title = {Last electroweak WIMP standing: pseudo-dirac higgsino status and compact stars as future probes},
volume = {42},
journal = {Chinese Physics C},
doi = {10.1088/1674-1137/42/4/043105}
}

@article{Leane:2023woh,
    author = "Leane, Rebecca K. and Smirnov, Juri",
    title = "{Dark matter capture in celestial objects: treatment across kinematic and interaction regimes}",
    eprint = "2309.00669",
    archivePrefix = "arXiv",
    primaryClass = "hep-ph",
    reportNumber = "SLAC-PUB-17729, LTH-1347",
    doi = "10.1088/1475-7516/2023/12/040",
    journal = "JCAP",
    volume = "12",
    pages = "040",
    year = "2023"
}

@article{Besla:2019xbx,
    author = "Besla, Gurtina and Peter, Annika and Garavito-Camargo, Nicolas",
    title = "{The highest-speed local dark matter particles come from the Large Magellanic Cloud}",
    eprint = "1909.04140",
    archivePrefix = "arXiv",
    primaryClass = "astro-ph.GA",
    doi = "10.1088/1475-7516/2019/11/013",
    journal = "JCAP",
    volume = "11",
    pages = "013",
    year = "2019"
}

@article{Reynoso-Cordova:2024xqz,
    author = "Reynoso-Cordova, Javier and Bozorgnia, Nassim and Piro, Marie-C{\'e}cile",
    title = "{The Large Magellanic Cloud: expanding the low-mass parameter space of dark matter direct detection}",
    eprint = "2409.09119",
    archivePrefix = "arXiv",
    primaryClass = "hep-ph",
    doi = "10.1088/1475-7516/2024/12/037",
    journal = "JCAP",
    volume = "12",
    pages = "037",
    year = "2024"
}

@article{Smith-Orlik:2023kyl,
    author = "Smith-Orlik, Adam and others",
    title = "{The impact of the Large Magellanic Cloud on dark matter direct detection signals}",
    eprint = "2302.04281",
    archivePrefix = "arXiv",
    primaryClass = "astro-ph.GA",
    doi = "10.1088/1475-7516/2023/10/070",
    journal = "JCAP",
    volume = "10",
    pages = "070",
    year = "2023"
}

@article{Patel:2015tea,
    author = "Patel, Hiren H.",
    title = "{Package-X: A Mathematica package for the analytic calculation of one-loop integrals}",
    eprint = "1503.01469",
    archivePrefix = "arXiv",
    primaryClass = "hep-ph",
    doi = "10.1016/j.cpc.2015.08.017",
    journal = "Comput. Phys. Commun.",
    volume = "197",
    pages = "276--290",
    year = "2015"
}

@article{Patel:2016fam,
    author = "Patel, Hiren H.",
    title = "{Package-X 2.0: A Mathematica package for the analytic calculation of one-loop integrals}",
    eprint = "1612.00009",
    archivePrefix = "arXiv",
    primaryClass = "hep-ph",
    doi = "10.1016/j.cpc.2017.04.015",
    journal = "Comput. Phys. Commun.",
    volume = "218",
    pages = "66--70",
    year = "2017"
}

\end{document}